\documentclass[useAMS,usenatbib]{mn2e}

\usepackage[dvips]{graphicx}
\usepackage{graphics}
\usepackage{psfig}
\usepackage{lscape}
\usepackage{amssymb}
\usepackage{amsmath}
\input{epsf}
\usepackage{booktabs}
\usepackage{subfigure}
\usepackage{graphicx}
\usepackage{wrapfig}
\usepackage{natbib}
\usepackage{color,soul}
\def\wl{\par \vspace{\baselineskip}}

\graphicspath{{Figures/}}

\title[Spectral Variability in H0557-385]{Absorption at the dust sublimation radius and the dichotomy between X-ray and optical 
classification in the Seyfert galaxy H0557-385\thanks{Based on observations obtained at the Southern Astrophysical Research (SOAR) telescope, which is a joint project of the Minist\'{e}rio da Ci\^{e}ncia, Tecnologia, e Inova\c{c}\~{a}o (MCTI) da Rep\'{u}blica Federativa do Brasil, the U.S. National Optical Astronomy Observatory (NOAO), the University of North Carolina at Chapel Hill (UNC), and Michigan State University (MSU).}}
\author[D. Coffey et al.]{D. Coffey$^{1,2}$\thanks{E-mail:
coffeydg@tcd.ie}, A. L. Longinotti$^{1,3}$, A. Rodr\'{i}guez-Ardila$^{4}$, M. Guainazzi$^{1}$, G. Miniutti$^{5}$,
\and S. Bianchi$^{6}$, I. de la Calle$^{1}$, E. Piconcelli$^{7}$, L. Ballo$^{8}$, M. Linares$^{9,10}$
\\
$^{1}$ XMM-Newton Science Operation Centre, RSSD-ESA, ESAC, P.O. Box 78, 28691 Villanueva de la Ca{\~n}ada, Madrid, Spain\\
$^{2}$ School of Physics, Trinity College Dublin, Dublin 2, Ireland\\
$^{3}$ MIT Kavli Institute for Astrophysics and Space Research, 77 Massachusetts Avenue, Cambridge, MA 02139, USA \\
$^{4}$ Laboratorio Nacional de Astrofi­sica Itajuba - MG, Brazil \\ 
$^{5}$ Centro de Astrobiolog\'ia (CSIC-INTA), Dep. de Astrof\'isica; ESAC, PO Box 78, Villanueva de la Ca\~nada, E-28691 Madrid, Spain \\
$^{6}$ Dipartimento di Matematica e Fisica, Universit\`a degli Studi Roma Tre, Via della Vasca Navale 84, I-00146, Rome, Italy \\  
$^{7}$ Osservatorio Astronomico di Roma (INAF), via di Frascati 33, 00040 Monte Porzio Catone (Roma) Italy \\
$^{8}$ INAF - Osservatorio Astronomico di Brera, via Brera 28, I-20121 Milan, Italy \\
$^{9}$ Instituto de Astrof{\'i}sica de Canarias, c/ V{\'i}a L{\'a}ctea s/n, E-38205 La Laguna, Tenerife, Spain \\ 
$^{10}$ Universidad de La Laguna, Dept. Astrof{\'i}sica, E-38206 La Laguna, Tenerife, Spain \\
}

\begin{document}

\date{}

\pagerange{\pageref{firstpage}--\pageref{lastpage}} \pubyear{2014}

\maketitle

\label{firstpage}

\begin{abstract}

\noindent In this work, the analysis of multi-epoch (1995-2010) X-ray observations of the Seyfert 1
galaxy H0557-385 is presented. The wealth of data presented in this analysis show that the source exhibits dramatic spectral variability, from a typical unabsorbed 
Seyfert 1 type spectrum to a Compton-thin absorbed state, on time scales of $\sim$\,5 years. This extreme change in spectral shape can be attributed to variations in 
the column density and covering fraction of a neutral absorbing medium attenuating the emission from the central continuum source. Evidence for Compton reflection of the 
intrinsic nuclear emission is present in each of the spectra, though this feature is most prominent in the low-state spectra, where the associated Fe emission line 
complex is clearly visible. 
In addition to the variable absorbing medium, a warm absorber component has been detected in each spectral state. 
Optical spectroscopy concurrent with the 2010 {\it{XMM-Newton}} observation campaign have detected the presence of broad optical emission 
lines during an X-ray absorption event. 
From the analysis of both X-ray and optical spectroscopic data, it has been inferred that the X-ray spectral variability is a 
result of obscuration of the central emission region by a clumpy absorber covering~$\ge$\,80 per cent of the source 
with an average column density of N$_\text{H}$\,$\sim$\,7$\times$10$^{23}$ cm$^{-2}$, and which is located outside the broad line region at a distance from the central 
source consistent with the dust sublimation radius of the AGN.

\end{abstract}

\section{Introduction}
\label{sec:intro_335}

This work documents the X-ray and optical spectral analysis of the extreme ``changing-look'' Seyfert 1 active galactic nucleus (AGN) H0557-385.
AGN that exhibit significant X-ray absorption variability have been the focus of intensive research efforts in recent years, 
since the interpretation of X-ray absorption variability patterns allow the innermost regions of such objects to be studied.   
In some cases, where the data permitted, the X-ray absorbing medium has been inferred to exist on scales coincident with the 
optical Broad Line Region \citep[BLR, e.g.][]{risaliti07,bianchi09,sanfrutos13}, which is often characterised by short time scale (days/weeks) variability.
On the other hand, long time scale (months/years) spectral variations have been associated with absorption by material present in the circumnuclear torus 
\citep[e.g.][]{piconcelli07,rivers11,miniutti14}, first inferred to exist as the main component of the Unification Model for AGN \citep{antonucci93,urry95}, and eventually spatially resolved via interferometric 
mid-infrared observations \citep[e.g. NGC 1068,][]{jaffe04}.

In order to fully explain the observed properties of AGN, theoretical considerations have shown that the circumnuclear torus is required to consist of a distribution 
of discrete, or clumpy, clouds at distances from the supermassive black hole (SMBH) of not more than a few pc 
\citep{krolik88,elitzur06,nenkova08a, nenkova08b}. 
Observational evidence in favour of a clumpy toroidal absorber has been obtained via interferometry in the 8\,-\,13 $\mu$m range in the case of the Circinus 
Galaxy \citep{tristram07}. 
In addition to the associated obscuration effects, a clumpy circumnuclear torus would 
imply that the Seyfert 1/2 dichotomy is not only dependent on the torus orientation angle with respect to the observer,
but is also dependent on the probability of the observers line of sight (LOS) intercepting a toroidal cloud. Therefore it would be expected that,
assuming a clumpy torus, the probability of directly observing the AGN continuum source would always be finite \citep{elitzur08,elitzur12}.

H0557-385 (z=0.03387) was originally identified as a Seyfert 1 AGN by \citet{fairall82}.
An early analysis into the nature of this source was based on data obtained by {\it{ASCA}}, and was presented by \citet{turner96}, 
where the authors report the presence of continuum absorption below 2 keV.
Similar absorption features were  detected in subsequent {\it{XMM-Newton}} observations presented by \citet[][hereafter A06]{ashton06}, where the spectra suggested that 
line of sight absorption was present in two distinct ionization phases. 
Here a phase is defined as gas at a particular column density and ionisation 
parameter\footnote{The ionisation parameter is defined as $\xi$\,=\,L/nr$^{2}$, where L is the 1\,-\,1000 Rydberg ionising luminosity (erg s$^\text{-1}$),
n is the gas density (cm$^\text{-3}$), and r is the distance from the ionising source to the absorbing gas (cm$^\text{-2}$)}.
A06 also report the presence of neutral absorption (N$_\text{H}$\,=\,1.2$\times$10$^{21}$ cm$^{-2}$) attenuating the primary emission, however the data did not allow 
the location of this component to be well constrained. Further {\it{XMM-Newton}} observations carried out in 2006 showed a dramatic change in spectral shape, 
with a decrease in flux by a factor of $\sim$10. \citet[][hereafter L09]{longinotti09} interpret this change in spectral shape as being due to a partial covering 
of the source by neutral circumnuclear clouds. This paper presents the outcome of an intensive observational campaign that was launched in 2010, which includes 
{\it{XMM-Newton}} and {\it{Swift}} observations in the X-ray domain, and optical spectroscopy obtained at the 4.1 m Southern Astrophysical Research Telescope (SOAR, Chile). 
Archival X-ray data are also included in this work. Following from the work of L09, this analysis re-examines the spectral variability in terms of the partial covering 
scenario, using the multi-epoch observational data to impose constraints on the geometry, location, and composition of the absorbing structures.

This paper is organised as follows: In Section~\ref{obs}, the acquisition of all X-ray and optical data used in this analysis will be documented. 
Section~\ref{analysis} will outline the theoretical model that has been developed, as well as a statistical investigation of the assumptions made while modelling the 
spectra. Section~\ref{sec:disc} will then present the physical interpretations of the spectral results, 
as well as a discussion of the nature of the absorbing medium in H0557-385.
The adopted cosmological parameters are H$_{0}$\,=\,70 km s$^{-1}$ Mpc$^{-1}$, and $\Lambda$\,=\,0.73.

\section{Observations and Data reduction}\label{obs}

\subsection{X-ray Observations}

\begin{table*}      
\centering  
\caption{\label{tab:log}Observation log of H0557-385. For observations 1-9, the fluxes have been measured by applying the best fit model to each individual data set.
For the \textit{Swift} observations, the flux was measured by fitting a powerlaw to each individual data set.\protect\\
$^{1}$ Effective exposure time, or the net exposure time after all filtering processes have been carried out.}
\begin{tabular}{l c c c c c c c }      
\toprule 
\midrule

\multicolumn{1}{c}{Obs. $\#$}
& \multicolumn{1}{c}{Mission}
& \multicolumn{1}{c}{Obs. Id.}
& \multicolumn{1}{c}{Date}
& \multicolumn{1}{c}{Eff. Exp.$^{1}$}
& \multicolumn{1}{c}{Flux$_{0.3-2}$} 
& \multicolumn{1}{c}{Flux$_{2-10}$}\\

&  &  & year-month-day & (ks) & 10$^{-12}$ erg cm$^{-2}$ s$^{-1}$ & 10$^{-12}$ erg cm$^{-2}$ s$^{-1}$ \\ 

\midrule

1 & ASCA & 73070000 & 1995-03-23 & 37 & 1.5$^{+1.0}_{-0.8}$ & 21.8$^{+1.5}_{-3}$  \\
2 & BeppoSAX & 511090031 & 2001-01-26 & 19 & 9.6$\pm$0.9 & 34.3$\pm$1.6 & \\
3 & XMM & 0109130501 & 2002-04-04 & 3 & 10.3$^{+0.3}_{-0.5}$ & 34.4$^{+0.7}_{-0.8}$ \\
4 & XMM & 0109131001 & 2002-09-17 & 4 & 10.93$^{+0.2}_{-0.11}$ & 42.5$\pm{0.7}$ \\
5 & XMM & 0404260101 & 2006-08-11 & 41 & 0.43$^{+0.07}_{-0.02}$ & 3.48$^{+0.06}_{-0.2}$ \\
6 & XMM & 0404260301 & 2006-11-03 & 56 & 0.37$^{+0.08}_{-0.03}$ & 3.02$^{+0.09}_{-0.18}$ \\
7 & XMM & 0651530201 & 2010-10-15 & 21 & 0.60$^{+0.02}_{-0.15}$ & 3.3$^{+0.1}_{-1.1}$ \\
8 & XMM & 0651530301 & 2010-10-19 & 24 & 0.56$^{+0.08}_{-0.09}$ & 3.0$^{+0.1}_{-0.3}$ \\
9 & XMM & 0651530401 & 2010-10-31 & 21 & 0.52$^{+0.08}_{-0.09}$ & 2.93$^{+0.05}_{-0.4}$ \\

\midrule

\multicolumn{7}{c}{\textit{Swift} Monitoring} \\
\midrule
\multicolumn{1}{c}{} &
\multicolumn{1}{c}{} &
\multicolumn{1}{c}{Obs. Id.} &
\multicolumn{1}{c}{Date} &
\multicolumn{1}{c}{Eff. Exp.$^{1}$} &
\multicolumn{1}{c}{Flux$_\text{0.4-2}$} &
\multicolumn{1}{c}{Flux$_\text{2-5}$}
\\
& &  & year-month-day & (ks) & 10$^\text{-13}$\,erg\,cm$^\text{-2}$\,s$^\text{-1}$ & 10$^\text{-13}$\,erg\,cm$^\text{-2}$\,s$^\text{-1}$  \\
\midrule 

& &00090392001   &  2010-04-03    &     3.5 & 4.7$_{-1.9}^{+1.6}$  & 6.9$_{-2.5}^{+2.2}$     \\
& & 00090392002   &  2010-04-23    &     3.4 & 4.6$_{-1.6}^{+1.9}$  & 6.7$_{-2.4}^{+2.1}$     \\
& & 00090392003   &  2010-05-13    &     3.4 & 4.6$_{-1.4}^{+1.2}$  & 6.6$_{-2.4}^{+2.2}$     \\
& & 00090392004   &  2010-06-02    &     2.6 & 4.8$_{-1.7}^{+1.7}$  & 7.0$_{-2.5}^{+2.5}$     \\
& & 00090392005   &  2010-06-22    &     3.4 & 5.6$_{-2.1}^{+1.8}$  & 8.4$_{-2.8}^{+2.4}$     \\
& & 00090392006   &  2010-07-12    &     4.2 & 5.1$_{-1.4}^{+1.6}$  & 7.7$_{-1.7}^{+2.1}$ 	\\
& & 00090392007   &  2010-08-01    &     3.7 & 3.9$_{-1.8}^{+1.2}$  & 5.6$_{-1.6}^{+1.7}$ 	\\
& & 00090392008   &  2010-08-21    &     4.1 & 3.9$_{-1.4}^{+1.4}$  & 5.8$_{-2.0}^{+1.9}$ 	\\
& & 00090392009   &  2010-09-10    &     2.6 & 7.1$_{-2.5}^{+2.2}$  & 10.3$_{-4.1}^{+3.6}$	\\
& & 00090392010   &  2010-09-14    &     1.6 & 6.8$_{-3.8}^{+3.6}$  & 9.8$_{-4.4}^{+4.2}$ 	 \\
& & 00090392011   &  2010-09-30    &     3.4 & 4.6$_{-1.8}^{+1.8}$  & 6.8$_{-2.8}^{+2.0}$ 	 \\
& & 00090392012   &  2010-10-20    &     3.6 & 5.6$_{-2.0}^{+1.7}$  & 8.2$_{-2.6}^{+2.6}$ 	 \\
& & 00090392013   &  2010-11-09    &     3.8 & 5.8$_{-1.6}^{+1.5}$  & 8.3$_{-2.4}^{+3.1}$ 	 \\
& & 00090392014   &  2010-11-29    &     3.8 & 5.9$_{-1.4}^{+1.8}$  & 8.7$_{-2.1}^{+2.6}$ 	\\
& & 00090392015   &  2010-12-19    &     2.3 & 5.5$_{-1.6}^{+2.2}$  & 7.8$_{-2.8}^{+3.2}$   \\
& & 00090392016   &  2011-01-08    &     3.0 & 4.1$_{-1.6}^{+2.0}$  & 6.4$_{-2.4}^{+2.5}$   \\
& & 00090392017   &  2011-01-28    &     3.6 & 4.6$_{-2.0}^{+1.4}$  & 6.5$_{-2.4}^{+2.6}$   \\
& & 00090392018   &  2011-02-17    &     3.9 & 4.4$_{-1.4}^{+1.6}$  & 6.5$_{-1.8}^{+2.2}$   \\
& & 00090392019   &  2011-03-09    &     3.7 & 5.5$_{-1.4}^{+2.1}$  & 8.3$_{-2.5}^{+2.4}$   \\
& & 00091016001   &  2011-06-28    &     1.6 & 3.4$_{-2.4}^{+2.0}$  & 5.0$_{-2.7}^{+3.7}$   \\
& & 00091016002   &  2011-06-29    &     2.3 & 4.3$_{-1.8}^{+2.0}$  & 6.2$_{-2.8}^{+2.6}$   \\
& & 00091016003   &  2011-08-11    &     3.7 & 3.3$_{-1.5}^{+1.5}$  & 4.9$_{-2.4}^{+2.2}$   \\
& & 00091016004   &  2011-09-24    &     3.2 & 4.8$_{-2.0}^{+1.5}$  & 6.9$_{-3.2}^{+2.1}$   \\
& & 00091016005   &  2011-11-07    &     2.0 & 2.8$_{-2.1}^{+1.7}$  & 4.0$_{-2.3}^{+1.9}$   \\
& & 00091016006   &  2011-11-10    &     1.5 & 4.1$_{-3.1}^{+2.8}$  & 5.9$_{-3.0}^{+3.9}$   \\
\midrule
\bottomrule \\
\end{tabular}
\end{table*}

To date, there have been  seven {\it{XMM-Newton}} \citep{jansen01} observations (see Table~\ref{tab:log}) of the Seyfert 1 AGN H0557-385 taken between 2002 and 2010. 
In addition, Table~\ref{tab:log} includes {\it{BeppoSAX}} and {\it{ASCA}} observations. Spectral products for both observatories were downloaded from 
the {\it{BeppoSAX}} archive interface and the Tartarus data base, respectively.
The {\it{BeppoSAX}} observations of this source have been presented by \citet{quadrelli03} and \citet{dadina07}. 
The EPIC pn CCD operated in large window mode for the 2002 observations, 
and in small window mode for all other observations. Both MOS CCDs operated in small window mode for each observation.
The observation data files have been processed using the {\it{XMM-Newton}} \textsc{science analysis systems} (SAS) version 13.0.0, 
including the latest calibration files available as of June 2013.  

Intervals of high background activity were removed by first extracting a light curve in the energy range 10\,-\,12 keV, 
and then applying rate thresholds of 0.35 counts s$^{-1}$ and 0.4 counts s$^{-1}$ for the EPIC MOS and EPIC pn respectively. 
All observations showed relatively low levels of background contamination, except for Obs. 0404260101, 
where the event list is a combination of two separate event lists that were generated during the observation (L09). 

Data obtained from the EPIC pn camera will be the primary focus of the following analysis, and for this reason, 
the description of all subsequent data reduction and spectral fitting procedures will refer only to EPIC pn spectra, unless otherwise stated. 
Circular source and background regions were extracted using \textsc{pattern} 0\,-\,4. For each of the 2006 and 2010 observations, 
the 0.3\,-\,10 keV count rate was below the critical value (small window mode: 25 counts s$^{-1}$) for which pile-up occurs, 
while for the 2002 observations the SAS task \textsc{epatplot} was run to verify that the spectra were not affected by pile-up. 
Inspection of the light curves for each observation reveals that there is no significant spectral variation on the short time scales 
over which the observations were carried out, therefore the full time-integrated spectra will be used in the following analysis for each observation.

For all observations RGS spectral products were obtained by running the SAS task \textsc{rgsproc}. 
However, the signal to noise ratio of the 2006\,-\,2010 data is so low that no accurate spectral measurements could be made. 
The RGS data of the 2002 observations have been presented in great detail in a previous analysis of this source (A06). 
Since a detailed analysis of the warm absorber is out of the scope of this paper, their results will be assumed and quoted when necessary. 
For these reasons, the RGS data will not be included in the present analysis.

From 2010 April to 2011 November the {\it{Swift}} observatory monitored H0557-385 every 3\,-\,4 weeks (see Table~\ref{tab:log}). 
The data from the X-ray Telescope (XRT, \citet{burrows05}) were always acquired in Photon Counting mode. 
The data were reduced with \textsc{xrtpipeline} 0.12.6, spectral products were extracted with \textsc{xselect} from source and 
background circular regions of 25 and 35 arcsec of radius, respectively.

\subsection{Optical Observations}

This source was also observed using the {\it{SOAR/Goodman}} optical spectrograph during the period 
November 2010\,-\,January 2011. The observations were carried out using the 600 l/mm grating and the 0.8'' slit width oriented at the 
position angle PA\,=\,0$^{\circ}$. The detector consisted of a Fairchild CCD 2048$\times$2048 pixels with 
a spatial scale of 0.15''/pixel. This setup provides a spectral resolution of 3.2~\AA\ 
and a spectral coverage from 4360~\AA\ to 6950~\AA, allowing the simultaneous detection of 
H$\gamma$, H$\beta$, [O\,{\sc iii}]~$\lambda$5007, and H$\alpha$. The first observation was 
carried out on 2010 October 17 (UT), consuming a total of 1.7 h; the second on 2010 November 3 (UT), 
with a total of 2.3 h; and the last on 2011 January 30 (UT), lasting 2 h. The former two 
visits were nearly simultaneous with {\it{XMM-Newton}} pointings, on 2010 October 15 and 2010 October 31. In all cases, 
the seeing was ~0.8'' with no photometric conditions.

The observing procedure consisted of recording three individual science frames of 900 s of 
integration each, totaling 2700 s on-source. Right before or after, a lamp frame of CuHeAr 
was obtained for wavelength calibration, followed by the observation of a spectroscopic 
standard star for flux calibration and deriving instrument response. Flats and bias frames 
were taken as diurnal calibrations. 

The optical data were reduced following standard IRAF procedures, that is, bias 
subtraction and division by a normalised flat field spectrum. The spectra were then 
extracted by summing up the signal recorded for the galaxy along the spatial direction. 
Individual 1D spectra of each night were wavelength calibrated using CuHeAr lamp frames 
and then combined to obtain a single 1D spectrum. It was then flux calibrated using the 
spectrum of the standard star LTT~2415, for the observations of October and November, and 
Hiltner~600 for that of January.

\begin{figure}
	\begin{center}
  	\psfig{figure=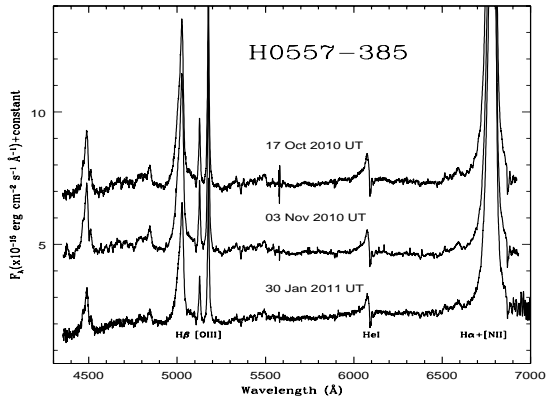,width=8cm,height=6cm}
	\caption{\label{fig:H0557-opt}Spectra of H0557-385 taken on the three different 
visits to this source. The observations of 2010 Oct 17 and 2010 Nov 03 were 
displaced along the vertical direction for displaying purposes.}
 	\end{center}
\end{figure}

Finally, it was necessary to correct the data for Galactic extinction.
Here, a value of E(B-V)\,=\,0.038 was adopted, as determined by \citet{schlafly11}.
Figure ~\ref{fig:H0557-opt} shows the wavelength and flux calibrated optical
spectra taken on the three different dates: 2010-10-17, 2010-11-03,  and
2011-01-30. Overall, the agreement in the continuum shape and  emission
line features is very good. Differences in the continuum  level, mostly
redward of 5300~\AA, can be seen but these are of
the order of 10 per cent or less. Considering that the observations  were
carried out under non-photometric conditions and the slit was  not
oriented at the parallactic angle, differences in the flux
level between the different spectra are expected, and can account  for the
apparent variability.

{\it{XMM-Newton's}} Optical Monitor (OM) telescope carried out observations 
of the source simultaneously with the EPIC observations. 
All OM data has been processed using the standard SAS task \textsc{omichain}. For the 2002 and 2006 observations, 
the OM was operated with the following optical and UV filters: 5430 (V), 4500 (B), 3440 (U), 2910 (UVW1), 2120 (UVW2), and 2310 (UVM2) \AA. 
For the 2010 observations only the UV filters were used: 3440 (U), 2910 (UVW1), and 2310 (UVM2) \AA.

\section{Spectral Analysis}\label{analysis}

The following analysis was carried out using \textsc{xspec} version 12.8.0f \citep{xspec96}. Errors are quoted throughout at the 90 per cent confidence level 
($\Delta\chi^{2}$=2.71) for a single parameter of interest. For each of the following spectral models, 
a Galactic column density of 4$\times$10$^{20}$ cm$^{-2}$ \citep{dickey90} is applied, and all elemental abundances are set to solar values unless stated otherwise.

\begin{figure}
	\centering
		\includegraphics[height=8cm,width=6cm,angle=270]{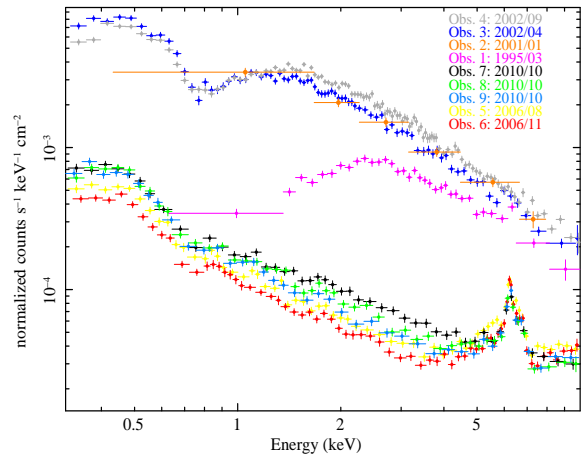}
	\caption{\label{fig:spectra}\textit{XMM-Newton, ASCA}, and \textit{BeppoSAX} observations of H0557-385.}
\end{figure}

The X-ray spectra for the nine deepest observations listed in Table~\ref{tab:log} are shown in Figure~\ref{fig:spectra}. 
It is noted here that the detailed spectral analysis described in the following sections applies only to Obs. 1\,-\,9 
due to the superior signal to noise ratio of these spectra. 
The poorer signal to noise ratio of the \textit{Swift} data render it inappropriate for detailed spectral modelling, 
and so it will be analysed separately in Section~\ref{swift}.

To allow for clearer visual comparison, the spectra in Figure~\ref{fig:spectra} have been divided by the corresponding detector effective area.
From inspection of Figure~\ref{fig:spectra}, it can be seen that there is dramatic spectral variability observed in this source across the entire 0.3\,-\,10 keV band.
Three main spectral states have been identified in this source; an extreme change in spectral shape occurs between the 
intermediate- ({\it{ASCA}} 1995), high- ({\it{XMM-Newton}} 2002),  and low- ({\it{XMM-Newton}} 2006/2010) state spectra, on a time scale of years. 
In addition, a less dramatic variation can be seen among the low-state spectra, which occurs on shorter time scales of weeks/months.

The overall strategy for this analysis is as follows; first a model that explains the low-state variability will be defined. 
Once a robust best-fit for the low-state spectra is established, it will then be used as a global model and extended to all other spectral states. 
In this way it will be tested if the same physical processes can explain both phases of the variability.

\subsection{Determining a Global Spectral Model}
\label{sec:global model}

The spectral analysis of the low-state spectra began with an analysis of the {\it{XMM-Newton}} Obs. 6 and 7 (see Table~\ref{tab:log}), since they represent 
the upper and lower bounds to the observed low-state variability.
Since the presence of a two-phase warm absorber in this source is already well documented  in the literature (A06 and L09), this component 
was included in the baseline model that consists of an intrinsic X-ray power law absorbed by a two-phase warm absorber system. The warm absorber is 
approximated by two \textsc{zxipcf} components \citep{miller07}, and the Galactic column density by the \textsc{tbabs} model \citep{wilms00}.
A Gaussian line at 6.4 keV was added to fit the prominent 
Fe K$\alpha$ emission clearly visible in the spectra. Following from the conclusion in L09 that the variability between spectral states is due to a 
neutral partial covering, this component was added (\textsc{zpcfabs} in \textsc{xspec}) such that the intrinsic power law was then absorbed 
by both ionised and neutral absorption. 
The spectral indices of the primary emission were assumed to take a common value for both of the spectra 
(model components that are set to be equal between spectral states in this manner will be referred to throughout as ``tied'').
The column densities and ionisation parameters of the warm absorbers have also been tied for both epochs (a test of this assumption is presented in 
Section~\ref{sec:const}). The remaining parameters, which include the normalisation of the power law, and the column density and covering fraction of the neutral 
absorber, were left free to vary between the states.

This model, which can be considered the basis from which the global model will be built, does not provide a good fit to the data 
($\chi_{\nu}^{2}$\,=\,1.84), as is expected. From inspection of the residuals generated by this fit, it was seen that the model provided a 
poor fit to the soft X-ray spectrum. 
In particular, positive residuals were seen to be present around 0.9 keV, which were also observed in L09, where the feature was modelled with a Gaussian line 
with energy consistent with emission from the Ne\,\textsc{ix} triplet. 
Contributions to the 0.5\,-\,2 keV spectrum is generally expected in AGN from 
both emission from the photoionised narrow line region \citep[NLR,][]{bianchi06}, and
gas ionised by stellar activity in the 
host galaxy \citep[][]{lamassa12}.
To account for this in the CCD spectra, and following the approach presented in \citet{miniutti14}, 
the collisionally-ionised gas emission model \textsc{apec} was adopted as a phenomenological description of the residuals in H0557-385. 
Adding this component improved the fit considerably 
by $\Delta\chi^{2}$\,=\,175 for 2 degrees of freedom (DOF) for a plasma temperature of kT\,$\simeq$\,0.9 keV, 
with flux F$^\text{APEC}_\text{0.5\,-\,2}$\,=\,5.4\,$\pm$\,0.3$\times$10$^\text{-14}$ erg\,cm$^\text{-2}$\,s$^\text{-1}$. 
Since this component is associated in either case with extended regions, 
it is expected to remain constant over long time scales, and therefore the normalisation was tied for each epoch. 
In addition, because this emission is due to activity at larger scales, 
it is not absorbed by the neutral/ionised components that absorb the primary X-ray emission. 

\begin{table*}
\centering

\caption{\label{tab:multi_spec}Best-fitting parameters for the global fit to each spectral state. The tied components have been set to be equal for each individual 
data set, but are free to vary for the global fit. The variable components have been left free to vary for each individual data set. Apec plasma temperature in units of 
keV, norm in units of 10$^{-5}$ ph keV$^{-1}$ cm$^{-2}$ s$^{-1}$, \textsc{pexmon} norm in units of 10$^{-3}$ ph keV$^{-1}$ cm$^{-2}$ s$^{-1}$, column densities in units 
of 10$^{22}$ cm$^{-2}$, ionisation parameters, $\xi$, in units of erg cm s$^{-1}$, power law norm in units of 10$^{-2}$ ph keV$^{-1}$ cm$^{-2}$ s$^{-1}$.\protect\\
$^{1}$  Covering fractions for the fit to the high-state spectra are fixed to one, as it is not expected that this parameter is measureable in these unabsorbed states.\protect\\
$^{2}$ The unabsorbed 0.3 - 2 keV and 2 - 10 keV luminosities measured using the EPIC best-fit model in units of 10$^{44}$ erg s$^{-1}$.}

\wl

\begin{tabular}{l c c c c c c c c c} 
\toprule
\midrule

\multicolumn{9}{c}{Tied Components}
\\
\midrule

\multicolumn{2}{c}{APEC}
& \multicolumn{1}{c}{PEXMON}
& \multicolumn{2}{c}{Warm Absorber$_{1}$}
& \multicolumn{2}{c}{Warm Absorber$_{2}$}
& \multicolumn{1}{c}{Power Law}
& \multicolumn{1}{c}{Fit Statistic} 

 \\

 kT & Norm & Norm & N$_\text{H}$ & Log($\xi$) & N$_\text{H}$ & Log($\xi$) & $\Gamma$ & $\chi^{2}_{\nu}$ \\ 
\midrule 

0.853$^{+0.018}_{-0.02}$ & 1.94$\pm{0.09}$ & 2.78$\pm{0.11}$ & 2.65$\pm{0.04}$ & 2.11$\pm{0.01}$ & 0.243$\pm{0.004}$ & 0.27$\pm{0.01}$ & 1.978$\pm{0.004}$ & 1.35
\\

\midrule 
\midrule

\multicolumn{9}{c}{Variable Components} \\
\cmidrule{1-9}
\multicolumn{1}{c}{} &
\multicolumn{1}{c}{Obs. $\#$} &
\multicolumn{1}{c}{Power Law} &
\multicolumn{2}{c}{Neutral Absorber} &
\multicolumn{1}{c}{L$_{0.3-2}^{2}$} &
\multicolumn{1}{c}{L$_{2-10}^{2}$}
\\
& & Norm & N$_\text{H}$ & C$_\text{f}$ &  &  \\
\midrule 

& 1 & 1.01$\pm{0.03}$ & 2.27$\pm{0.08}$ & 0.917$\pm{0.013}$ & 0.5$\pm{0.3}$ & 0.7$\pm{0.1}$  \\

& 3 & 1.443$\pm{0.015}$ & 0.089$\pm{0.003}$ & 1$^{1}$ & 1.16$\pm{0.06}$ & 1.03$\pm{0.02}$ \\

& 4 & 1.769$\pm{0.014}$ & 0.138$\pm{0.003}$ & 1$^{1}$ & 1.417$\pm{0.014}$ & 1.27$\pm{0.03}$ \\

& 5 & 0.63$\pm{0.03}$ & 66.6$^{+1.6}_{-1.5}$ & 0.940$\pm{0.003}$ & 0.51$\pm{0.08}$ & 0.46$\pm{0.03}$ \\

& 6 & 0.64$^{+0.11}_{-0.08}$ & 76.2$^{+1.6}_{-1.5}$ & 0.952$^{+0.008}_{-0.007}$ & 0.51$\pm{0.11}$ & 0.47$\pm{0.03}$ \\

& 7 & 0.39$\pm{0.03}$ & 61$\pm{3}$ & 0.858$^{+0.011}_{-0.013}$ & 0.32$\pm{0.08}$ & 0.29$\pm{0.09}$ \\

& 8 & 0.49$\pm{0.04}$ & 78$\pm{3}$ & 0.90$\pm{0.01}$ & 0.40$\pm{0.06}$ & 0.37$\pm{0.04}$ \\

& 9 & 0.63$\pm{0.05}$ & 88$\pm{3}$ & 0.929$^{+0.006}_{-0.007}$ & 0.51$\pm{0.09}$ & 0.46$\pm{0.07}$ \\

\midrule
\bottomrule

 \end{tabular}

\end{table*}

The prominent Fe K$\alpha$ complex in the hard X-ray band is generally associated with emission from neutral gas, 
possibly located in the obscuring ``torus'' as suggested by the Unification Model for AGN \citep{antonucci93} \citep[see also][]{ghisellini94}.
Therefore, for physical consistency, the Gaussian line modelling the Fe K$\alpha$ emission was replaced with a neutral Compton reflection component \citep[\textsc{pexmon} 
in \textsc{xspec},][]{nandra07}. This component describes the Compton reflection of the primary emission by optically thick gas in a disk-like geometry around the source. 
The associated emission lines (Fe K$\alpha$ 6.4 keV, Fe K$\beta$ 7.05 keV, and Ni K$\alpha$ 7.47 keV) are included, as well as a Compton shoulder at 6.315 keV, 
the strength of which is dependent on the inclination angle, which is assumed here to be 45$^{\circ}$. The illuminating spectral index was set to be equal to that of the 
intrinsic power law, and the normalisation of this component was tied for both observations. This gives a slightly poorer fit ($\Delta\chi^{2}$\,=\,28 for the same DOF), 
however due to the physical relevance of the neutral 
reflection, and the fact that it still provides a relatively good fit, the analysis will proceed with the inclusion of this component. It is noted here that since the 
Compton reflection component is expected to exist on larger scales, it is not 
absorbed by the variable neutral absorber attenuating the primary X-ray emission.
This assumption will be validated in Section~\ref{sec:const}.

This model, which successfully accounts for the variability between the 2006 and 2010 observations, was then extended to the remaining low-state spectra. 
Again, for this fit, all parameters have been tied except for the normalisation of the intrinsic power law, and the column density and 
covering fraction of the neutral absorber. From visual inspection of the data-to-model ratio generated by this fit, it was clear that  
each of the low-state spectra are well explained by this model, 
validating the assumption that the variability observed in the low state spectra is due to neutral absorption attenuating the primary emission.

\begin{figure}
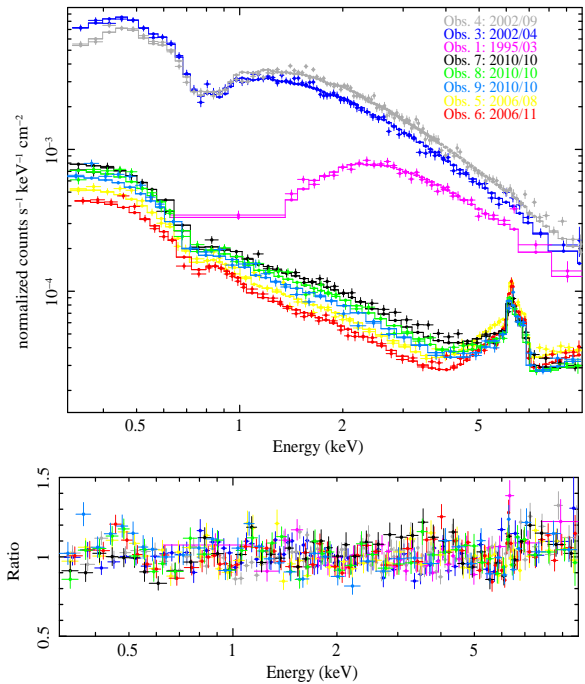

	\centering
	\begin{subfigure}
		\centering
		\includegraphics[height=8cm,width=6cm,angle=270]{ext_all_rat.ps}
	\end{subfigure}
	\begin{subfigure}
		\centering
		\includegraphics[height=8cm,width=3cm,angle=270]{ratio.ps}
	\end{subfigure}
	\caption{\label{fig:ratio} The upper panel shows the data and applied model for all of the spectra listed in Table~\ref{tab:multi_spec}, 
	while the lower panel gives the corresponding residuals for each fit.}
\end{figure}

To test if this holds for the intermediate- and high-state spectra, the model described above was then applied to the remaining {\it{XMM-Newton}} and {\it{ASCA}}
observations. The data, best-fit model, and data-to-model ratio are given in Figure~\ref{fig:ratio}, and the corresponding best-fit parameters and their 
errors are given in Table~\ref{tab:multi_spec}. Again it can be seen that the model provides a reasonable fit for the multi-epoch spectra, 
with a $\chi^{2}_{\nu}$\,=\,1.35, and no obvious structure present in the residuals (except for some residuals present around 6\,-\,7 keV, which are discussed
in Section~\ref{sec:physics_const}). 
Since the {\it{BeppoSAX}} spectrum is approximately coincident with the 2002 {\it{XMM-Newton}} spectra,
it was not included in the global fit, but instead was fit separately in the 0.4\,-\,8 keV range to test for consistency with the results quoted in 
Table~\ref{tab:multi_spec}. The model described above was fit to the {\it{BeppoSAX}} data, with all components tied to the best-fit values quoted in 
Table~\ref{tab:multi_spec}, except for the normalisation of the power law and column density of the neutral absorber, 
which were measured to be 1.44$\pm{0.08}\times$10$^\text{-2}$ ph keV$^\text{-1}$ cm$^\text{-2}$ s$^\text{-1}$ and 0.11$\pm{0.04}\times$10$^{22}$ cm$^{-2}$ respectively. 
As expected, these values are consistent with those found for Obs. 3. 
This fit gave a fit statistic of $\chi^{2}_{\nu}$\,=\,1.01, indicating that the model provides an extremely good fit to the {\it{BeppoSAX}} data.

Included in this model is a Gaussian line with energy fixed to 6.67 keV, 
accounting for the detection of the Fe\textsc{xxv} K$\alpha$ emission line
also reported in L09. 
This line is statistically required by some of the spectra in the global fit, 
but gave a measured normalisation consistent with zero in Obs 4, 8, and 9. 
Each of the spectra were fitted individually, and it was found that the 
upper limit on this line is consistent with the best fit value of the global model.
Therefore, in order to apply the global model to all of the spectra simultaneously, 
the normalisation of this line was kept fixed to its best fit value ($\simeq$\,3$\times$10$^\text{-6}$ ph cm$^\text{-2}$ s$^\text{-1}$) during the spectral fitting.

In the following sections, the assumptions made in modelling these spectra will be statistically investigated, 
while a physical interpretation of these assumptions will be provided in Section~\ref{sec:disc}.

\subsection{Power Law Normalisation and Neutral Absorber Covering Fraction Correlation}

\begin{figure*}
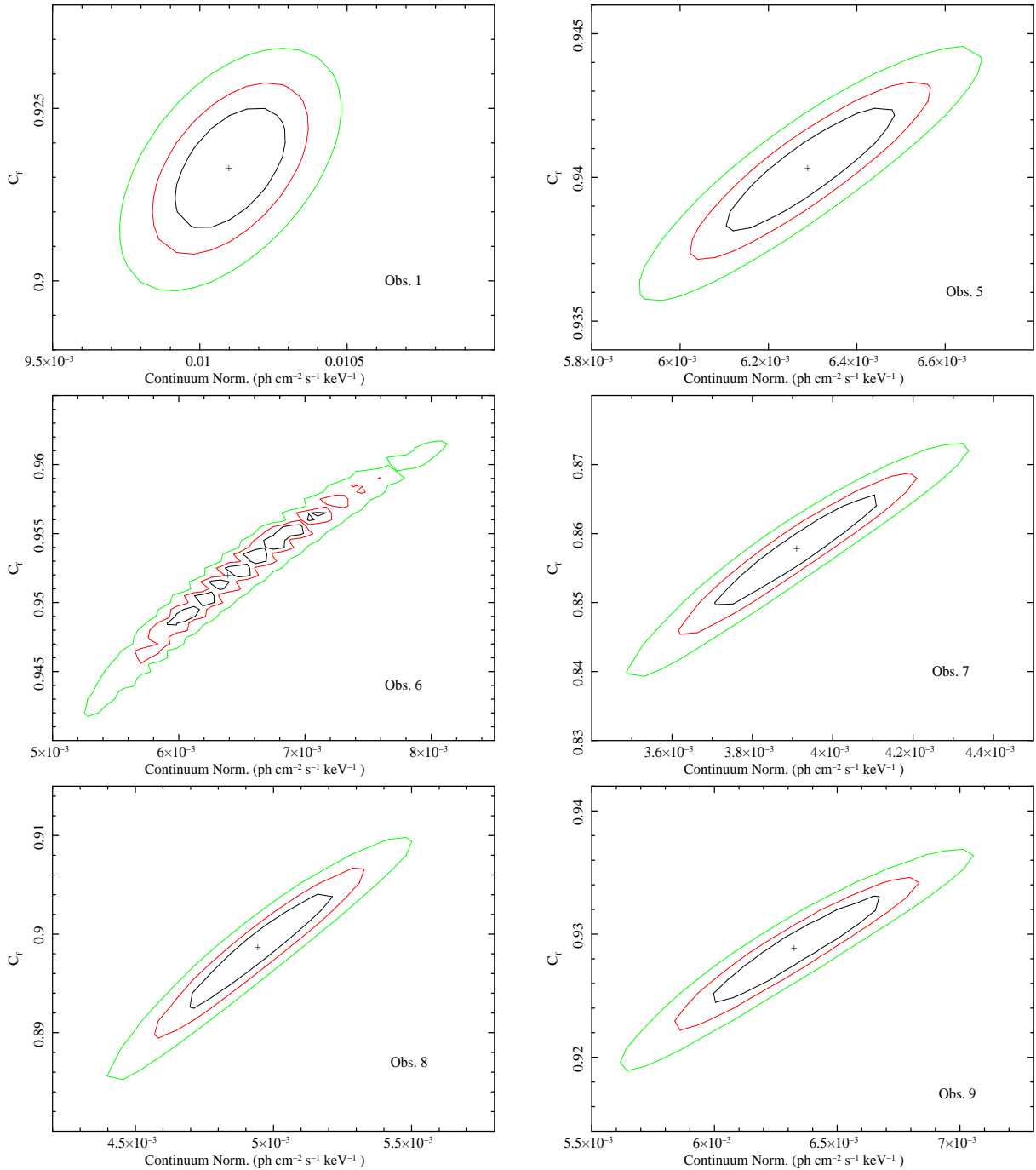

	\begin{tabular}{c c}
	 	\psfig{figure=./Figures/conf_cont/73070000_cf_norm.ps,width=8cm,height=6cm,angle=-90} &
		\psfig{figure=./Figures/conf_cont/0404260101_cf_norm.ps,width=8cm,height=6cm,angle=-90} \\
		\psfig{figure=./Figures/conf_cont/0404260301_cf_norm.ps,width=8cm,height=6cm,angle=-90} &
		\psfig{figure=./Figures/conf_cont/0651530201_cf_norm.ps,width=8cm,height=6cm,angle=-90} \\
		\psfig{figure=./Figures/conf_cont/0651530301.ps,width=8cm,height=6cm,angle=-90} &
		\psfig{figure=./Figures/conf_cont/0651530401_cf_norm.ps,width=8cm,height=6cm,angle=-90} \\

	\end{tabular}
\caption{\label{fig:conf_cont} The 68 (black), 90 (red) and 99 (green) per cent confidence contours for power law normalisation and covering fraction of the neutral 
absorber for {\it{ASCA}} Obs. 1 and {\it{XMM-Newton}} Obs. 5\,-\,9. The strong correlation between these two parameters is clearly present in Obs. 5\,-\,9. 
For Obs. 1 the effect is less pronounced since for these data the neutral absorber component is less prominent due to a lower column density 
(N$_\text{H}$\,$\sim$\,2$\times$10$^{22}$).}
\end{figure*}

For the large column densities measured by the model defined above, there is an intrinsic degeneracy between the normalisation of the powerlaw 
continuum and the covering fraction of the neutral absorber.
It was necessary therefore to take this effect into account when determining 
the true errors on these parameters for {\it{ASCA}} Obs. 1 and {\it{XMM-Newton}} Obs. 5\,-\,9 (where the covering fraction is $>$\,80 per cent in each case). 
To calculate the true errors on these parameters, the two dimensional confidence levels for these parameters were plotted 
for the 68, 90, and 99 per cent levels of confidence, which correspond to deviations in the $\chi^{2}$ best-fit value of 2.3, 4.61, and 9.21 respectively \citep{avni76}. 
The 90 per cent confidence level errors for these parameters can then be measured from the upper and lower bounds of the 90 per cent confidence contours. 
The errors quoted in Table~\ref{tab:multi_spec} for these parameters have been measured in this way.
The confidence contours for {\it{ASCA}} Obs. 1 and {\it{XMM-Newton}} Obs. 5-9 are shown in Figure~\ref{fig:conf_cont}.

\subsection{Tied Spectral Components}\label{sec:const}

When defining the global model, the ionised absorber components were tied to a common value for each of the spectra, 
assuming that they do not vary over the time scale of observation.
The ionised absorbers have little effect on the low-state spectra, 
where neutral absorption dominates, and therefore can only be reliably constrained by the high-state spectra. 
Using spectra 3 and 4, it is therefore possible to set a lower limit 
on the time scales over which the ionised absorbers remain constant. To do this, the model was again applied to the 2002 {\it{XMM-Newton}} spectra, 
this time also allowing the column densities and ionisation parameters of the ionised absorbers to vary for both spectra. The best-fit values for the 
ionised absorber components are given in Table~\ref{tab:warm_abs}. Allowing the ionised absorber components to be variable marginally improves the fit 
by $\Delta\chi^{2}$\,=\,29 for 4 DOF. However, from inspection of the parameters and their errors listed in Table~\ref{tab:warm_abs}, it can be seen that 
the parameters of the second warm absorber component are statistically consistent, while the ionisation parameter of the first warm absorber varies slightly 
between the two measurements. Therefore it can be concluded that these components are not observed to vary significantly between Obs. 3 and 4.

\begin{table}
\centering
\caption[''long'']{\label{tab:warm_abs} Best-fit parameters for the ionised absorber components when allowed to be free to vary for the 2002 high-state spectra. 
Column densities, N$_\text{H}$ in units of 10$^{22}$ cm$^{-2}$, ionisation parameters, $\xi$, in units of erg cm s$^{-1}$.}
 \begin{tabular}{l c c c c c}\\
\toprule
\midrule
\multicolumn{1}{c}{Obs. $\#$} &
\multicolumn{2}{c}{Warm Absorber$_{1}$} &
\multicolumn{2}{c}{Warm Absorber$_{2}$} \\

& N$_\text{H}$ & Log($\xi$) & N$_\text{H}$ & Log($\xi$)  \\
\midrule 

3 & 1.2$^{+0.3}_{-0.2}$ & 2.00$^{+0.16}_{-0.2}$ & 0.39$^{+0.10}_{-0.03}$ & 0.27$^{+0.08}_{-0.16}$  \\

4 & 0.9$\pm{0.3}$ & 1.54$^{+0.2}_{-0.19}$ & 0.39$^{+0.13}_{-0.05}$ & 0.27$^{+0.06}_{-0.2}$  \\

\midrule
\bottomrule
\end{tabular}
\end{table}   

The global spectral model also assumes that the spectral index of the power law is invariant between each of the spectra. To investigate if this assumption is valid, 
the spectral index was allowed to be free to vary for each observation in the global fit. 
The resultant best fit values were in the range 1.933\,-\,2.068. 
This lack of dramatic variability indicates that this component is adequately represented by a model that assumes 
an invariant spectral index. 

The reflection component (\textsc{pexmon} in \textsc{xspec}) was also assumed to remain constant over the time scale of observation.
To test this, the model was fit as described in Section~\ref{sec:global model}, but also allowing the normalisation of the \textsc{pexmon} component 
to be free to vary for each observation. For each of the low-state spectra and Obs. 4, the normalisation was found to be consistent with the
average value quoted in Table~\ref{tab:multi_spec}. For Obs. 1 and 3, the normalisation was found to be slightly higher than the average value
(6\,$\pm$\,2$\times$10$^{-3}$ ph keV$^{-1}$ cm$^{-2}$ s$^{-1}$), which may be linked to the higher intrinsic flux at these epochs. 
In order to test for obscuring material on the LOS to the reflection component, a neutral absorption component (\textsc{zpcfabs}) was applied to
\textsc{pexmon}. This model was fit with each parameter (except for the column density of the additional absorption component) 
tied to the best fit values given in Table~\ref{tab:multi_spec}. 
This fit gave column density values $\le$\,8$\times$10$^\text{20}$, consistent with the assumption that the \textsc{pexmon}
component is not observed through the same absorbing medium that obscures the primary emission.

\subsection{Variable Spectral Components}
\label{sec:variable_spec}

The model defined in section~\ref{sec:global model} proposes that the dramatic spectral variability is due primarily to variations in the neutral absorber 
obscuring the AGN primary emission. However, an alternate explanation for such dramatic spectral variability is that the emission from the active nucleus has faded, 
or switched off, leaving only reflected spectral imprints as indications of past activity \citep{guainazzi98,gilli00,matt03}. To test if this is the case in H0557-385, the global model was again applied 
to each of the low-state spectra, however this time the neutral absorption component was removed. This results in a very poor fit ($\chi^{2}_{\nu}$\,=\,5.42), 
demonstrating that the spectral variations in H0557-385 cannot be accounted for by only considering changes in the primary emission. In addition, it can be seen from 
Figure~\ref{fig:spectra} that emission at soft X-ray energies has increased in 2010 with respect to 2006, suggesting that nuclear emission is leaking through the absorber. 
This observation provides further evidence in favour of the partial covering model adopted in this analysis.

It was also assumed in section~\ref{sec:global model} that the intervening gas is composed of neutral material. 
The validity of this assumption was investigated by testing the effect of a variable absorber composed by ionised material.
This test was performed by replacing the neutral absorption component (\textsc{zpcfabs} in \textsc{xspec}) 
with an ionised absorption component (\textsc{zxipcf} in \textsc{xspec}). 
This model was fit to the low-state {\it{XMM-Newton}} data, with the following parameters left free to vary for each of the spectra: 
the column density, covering fraction, and ionisation parameter of the ionised absorber, and the normalisation of the intrinsic power law. 
Statistically, this model provides a better fit, with $\Delta\chi^{2}$\,=\,102 for 5 DOF when compared to 
the model where absorption is produced by neutral gas.
However, the inclusion of the ionised absorber yields a much higher value of the power law photon index i.e. $\Gamma$\,$\simeq$\,2.6. 
If the power law is forced to be the same as in the global model with neutral absorption (i.e. by fixing $\Gamma$\,$\simeq$\,2),  
a poorer fit is found ($\Delta\chi^{2}$\,=\,46 with an increase of 1 DOF). 

\begin{figure}
 \begin{center}
  \psfig{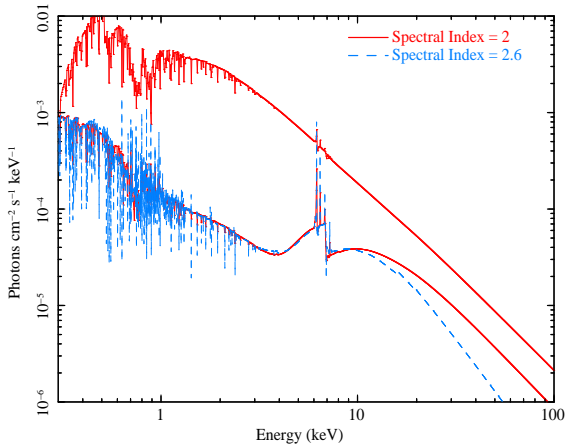}
  \caption{\label{fig:ion_abs} A comparison between the ionised and neutral absorption models is shown. The red lines show the high and low state neutral absorption models 
			       with $\Gamma$\,$\simeq$\,2. The dashed blue line represents the ionised absorption model which requires $\Gamma$\,=\,2.6. It is noted here that the 
			       intensities of the emission lines around 6 keV are not exactly aligned. This is due to the fact that the neutral absorption models 
			       (red lines) were fit using both high- and low-state spectra, while the ionised absorber model was fit using only the low-state spectra, 
			       thus giving slightly different values for the normalisation of the \textsc{pexmon} component.}
 \end{center}

\end{figure}

Figure~\ref{fig:ion_abs} shows that the change in the power law slope is probably the only observable that can be effectively used to gauge 
the ionization state of the variable absorber, because the ionised absorber imprints very mild absorption features on the continuum, 
which are virtually undetectable with CCD data at this flux. 
Nonetheless, the neutral and ionised absorber models present different spectral indices when fitted to the low-state data, 
as shown in the figure. This difference is mostly appreciable above 10 keV, 
a region of the spectrum that is not well sampled by current data. 
In fact, the data accumulated by the BAT instrument onboard {\it{Swift}} (in the energy range 15\,-\,150 keV) are integrated over a long time period, 
\citep[70 months,][]{baumgartner13}, therefore an absorber measured in the BAT spectrum might actually be the combined result of mixed spectral states. 
Therefore, the averaged BAT spectrum is not appropriate for constraining the 
ionised absorber model at high energies.

It is noted that the measured unabsorbed intrinsic power law spectral index of $\sim$\,2 is more in agreement with the mean value of the spectral index 
in the 3 - 10 keV range, $<\Gamma_{3-10}>$ = 1.91\,$\pm$\,{0.07}, found by \citet{nandra97}, 
by modelling a sample of 18 Seyfert 1 galaxies observed by {\it{ASCA}} and with the value $<\Gamma>$\,=\,1.89\,$\pm$\,{0.11} measured by \citet{piconcelli05}
in the 2 - 12 keV spectra of a sample of 40 QSO from the Palomar-Green (PG) Bright Quasar Survey sample. 
This remark provides further support to the assumption of a variable absorber composed of neutral material. 

Finally, it is also possible that the absorbing medium becomes ionised in response to the increase in intrinsic flux from the source during the high-state epoch.
This may lead to a decrease in the opacity of the absorbing medium, with the column density falling with increasing intrinsic flux \citep[e.g.][]{pounds04}.
However, this effect is not expected to occur in H0557-385, since the analysis carried out in A06 has shown that the 2002 RGS spectra require absorption by neutral
material.

\subsection{Results from the \textit{Swift} Monitoring}\label{swift}

\begin{figure}
 \begin{center}
  \psfig{figure=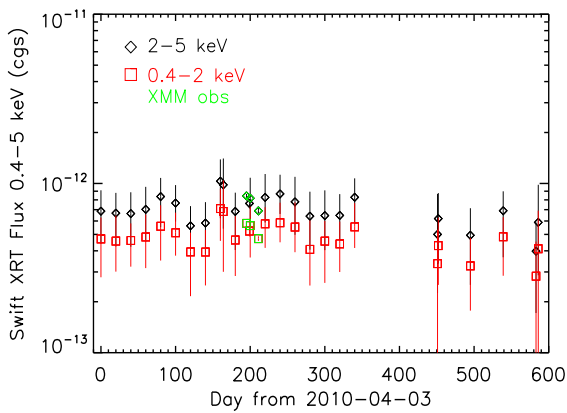,width=8cm,height=6cm}
\caption{\label{fig:flux_swift}Historic fluxes from the {\it{Swift}} XRT data obtained from 2010 March to 2011 November. For comparison, the figure includes also the three 
contemporaneous {\it{XMM-Newton}} observations of 2010 October. The three data points are all concentrated at Day 200 and can be recognized thanks to the smaller error bars 
compared to the {\it{Swift}} ones.}
 \end{center}

\end{figure}

Due to the low count rate, the \textit{Swift} spectra were rebinned with 5 counts per channel and the Cash statistic 
was applied for spectral fitting. Although each individual spectrum provides only basic information on the spectral shape of the source, it is very clear that {\it all} 
{\it{Swift}} snapshots caught H0557-385 still in the lowest flux state of 2006 and 2010.
Therefore, the flux was simply measured in the 0.4\,-\,2 keV and 2\,-\,5 keV bands by fitting a power law to the XRT spectra.
From the historic light curve presented in Figure~\ref{fig:flux_swift}, it is evident that during the {\it{Swift}} monitoring, which lasted approximately 1.5 years, 
the source does not undergo 
extreme variability (as between the states of 1995, 2002, and 2006). 
In addition, it is noted that the data obtained by the $\textit{Swift}$ UVOT show that no significant variability is detected at UV/optical wavelengths 
for the entire duration of the monitoring.
This information will be taken into account in the discussion presented  in Section~\ref{sec:disc}.

\subsection{Optical Spectroscopy of the BLR and Optical/UV photometry from \textbf{\textit{XMM-Newton}} Optical Monitor}\label{sec:opt_spec}

As the integrated emission line spectrum of H0557-385 is
likely affected by internal extinction and the observed continuum
includes contribution from the host galaxy, it is first necessary to
correct for these two effects before assessing if the AGN varied during
the different pointings.

Both internal extinction and contribution of stellar light were
determined using the code \textsc{starlight} \citep{cidfernandes04,cidfernandes05,mateus06,asari07}. 
Basically the code fits an observed spectrum O$_{\lambda}$ with a combination, in different proportions, of a number of simple stellar
populations (SSPs). In addition, in trying to describe the continuum  of a
Seyfert 1, the contribution of central engine should also be included.
Usually this component is represented by a featureless continuum
(FC; e.g. \citet{koski78}) of powerlaw form that follows the expression
F$_{\lambda}$\,$\propto$\,$\lambda^{\alpha}$. 
Therefore, this component was  also
added to the base of elements.

Extinction is modelled by \textsc{starlight} as due to foreground dust,
and parametrised by the V-band extinction A$_{\rm v}$. The extinction law of 
\citet{cardelli89} was used in this procedure
as it is widely used in AGNs.

The synthesis carried out with \textsc{starlight} shows that 30$\%$ of the
continuum emission at H$\beta$ is due to stellar population.
Also, it shows that the observed continuum is affected by
an A$_{\rm v}$ of 1.63, which translates into an internal
extinction E(B-V) of 0.53. This latter value is
in very good agreement with the E(B-V) of 0.54$\pm$0.08 derived
from the line flux ratio H$\alpha$/H$\beta$ measured in the
observed spectrum (adopting an intrinsic Balmer decrement of 3.1), giving
additional support to the spectral synthesis results.
Therefore, after correcting the three spectra for an E(B-V) of 0.53  and
subtracting the stellar population, we are left with the
continuum that can be attributed entirely to the AGN.

In order to determine if H0557-385 truly varied during the three visits, the 
method described in \citet{peterson91} was followed. It consists of using the strong, 
narrow [O\,{\sc iii}]~$\lambda$5007 line as an internal flux standard. The large spatial 
extend of the NLR and the low electron density (which implies in a 
very long recombination time) tend to damp out the effect of any short-term variation of 
the ionising continuum. Therefore, by making the well-justified assumption that the narrow 
line flux F$_{\rm [O\,{\textsc{iii}}]}$ is constant over the time scale of interest, variability in the 
continuum and the broad component of the H$\beta$ line can be discerned by measuring the ratios 
F$_{\lambda}$/F$_{\rm [O\,{\textsc{iii}}]}$ and  F$_{\rm H\beta_{BC}}$/F$_{\rm [O\,{\textsc{iii}}]}$, where
F$_{\lambda}$ refers to the continuum flux at some specific wavelength (here, taken to be 4870~\AA\  as in
\citet{peterson91}, in the laboratory rest frame) and F$_{\rm H\beta_{BC}}$ is the integrated flux
in the broad component of the H$\beta$ line. Because the spectra also included H$\gamma$, 
the ratio F$_{H\gamma_{BC}}$/F$_{\rm [O\,{\textsc{iii}}]}$ is also included as a consistency check, where 
F$_{H\gamma_{BC}}$ refers to the flux of the broad component of that line. It was decided that the H$\alpha$ line would not be used in this analysis because it is too 
far in wavelength from [O\,{\sc iii}]~$\lambda$5007,  making it susceptible to atmospheric dispersion effects, 
considering that the position angle during the observations was not aligned along the parallactic 
angle. 

Because the three spectra were taken using the same telescope and instrument setup, it is 
perfectly valid to measure the above line ratios for each date and then compare these
ratios among the different dates without the need of any normalisation. 
However, the observations of November 03 and January 30 have also been normalised to the [O\,{\sc iii}] flux. 

Table~\ref{tab:opt_data} lists the line fluxes measured in the three spectra. In order to
deblend the broad and narrow components from the observed profile, it was assumed that both H$\beta$ and
H$\gamma$ can be represented by a combination of Gaussian profiles. 
For these two H\,{\sc i} lines, two components were always necessary. For each of the three different dates, an excellent agreement was found in the number of components, 
full-width at half maximum (FWHM) values, and integrated emission line fluxes.

It was found that the narrow component of H$\beta$ has a FWHM of 910\,$\pm$\,40 km~s$^{-1}$ in 
velocity space. The broad component displayed a FWHM of 3140\,$\pm$\,90 km~s$^{-1}$. These
values are already corrected in quadrature by an instrumental broadening of 200 km~s$^{-1}$,
measured from the sky lines present in the spectra.

It is interesting to note that in all three spectra, the peak of the broad component of the 
Balmer lines is blueshifted relative to that of the narrow component by 11~\AA\ or 680\,km~s$^{-1}$.
The position of the narrow component of the H\,{\sc i} lines as well as that of 
[O\,{\sc iii}]~$\lambda$5007 coincides with the systemic velocity. These results indicate 
that part of the BLR in H0557-385 may be in an outflow. 

Table~\ref{tab:opt_data} also lists the line ratios derived from the optical observations. The lack 
of any variability between the three visits is evident, as the ratios measured 
in the spectra agree within the uncertainties (3$\sigma$). Therefore, it is concluded that the 
source remained stable during the sampled period.

\begin{table*}
\centering

\caption{\label{tab:opt_data}Optical emission line fluxes measured from the optical spectra of H0557-385.\protect\\
$^{a}$ In units of 10$^{-13}$\,erg\,cm$^{-2}$\,s$^{-1}$ \newline
$^{b}$Continuum flux at 4870~\AA. In units of  10$^{-13}$\,erg\,cm$^{-2}$\,s$^{-1}$\,\AA$^{-1}$}
\wl

\begin{tabular}{l c c c c} 

\toprule
\midrule

\multicolumn{1}{c}{}
& \multicolumn{1}{c}{2010-10-17}
& \multicolumn{1}{c}{2010-11-03}
& \multicolumn{1}{c}{2011-01-30} \\

\cmidrule{1-4}

Line Flux$^{a}$ & & & & \\

\cmidrule{1-4}

[O\,{\sc iii}] $\lambda$5007 & 5.02$\pm$0.23 & 5.2$\pm$0.19 & 3.48$\pm$0.15 \\

H$\beta_{NC}$  & 3.14$\pm$0.20 & 3.24$\pm$0.16 & 2.17$\pm$0.12  \\

H$\beta_{BC}$  & 10.51$\pm$0.71 & 10.9$\pm$0.53 & 7.25$\pm$0.41 \\

H$\gamma_{NC}$ & 1.64$\pm$0.24 & 1.74$\pm$0.28 & 1.14$\pm$0.27 \\

H$\gamma_{BC}$ & 3.14$\pm$0.72 & 3.98$\pm$0.77 & 2.79$\pm$0.95 \\

F$_{4870}^b$ & 0.11$\pm$0.02  & 0.12$\pm$0.03  & 0.08$\pm$0.02 \\

\cmidrule{1-4}

Flux Ratio & & & & \\

\cmidrule{1-4}

F$_{4870}$/F$_{[O\,{\textsc{iii}}]}\times$4870 \AA & 107$\pm$20 & 112$\pm$28 & 112$\pm$19 \\

F$_{H\beta}$/F$_{[O\,{\textsc{iii}}]}$ & 2.1$\pm$0.1 & 2.1$\pm$0.1 & 2.1$\pm$0.1 \\

F$_{\rm H\gamma}$/F$_{[O\,{\textsc{iii}}]}$ & 0.6$\pm$0.2 & 0.8$\pm$0.2 & 0.8$\pm$0.3 \\

\midrule 
\bottomrule
 \end{tabular}
\end{table*}

The ultraviolet flux densities have been checked in the M2 and W1 filters (2310 and 2910 \AA) from the OM telescope onboard {\it{XMM-Newton}}, which 
observed simultaneously to the X-ray instruments. These OM data are available for all {\it{XMM-Newton}} observations but one (2006 November). The OM data do not reveal 
significant variations in any band, therefore it is concluded that the variability of the X-ray emission observed over the years is not related to variation in the 
ultraviolet source.

\section{Discussion}\label{sec:disc}

The multi-epoch spectral model defined in Section~\ref{sec:global model} has shown that H0557-385 is a remarkable example of X-ray absorption variability. 
The emergent picture is that of an AGN in which most of the spectral features show no variation over the time scale of observation. Instead, the flux variability of 
this source can be attributed entirely to neutral material attenuating the AGN primary emission. Based on this, a physical discussion of the constant spectral components 
will first be provided, while an investigation into the nature of the X-ray absorber will be given in Sections~\ref{sec:structure} and~\ref{sec:origin}.

\subsection{Physical Interpretation of the Constant Spectral Components}\label{sec:physics_const}

The Fe K emission line complex that dominates the low-state spectra is also detected in the intermediate- and high-state spectra, though its shape is less pronounced due 
to the higher intrinsic flux. 
This feature has been modelled using a Compton reflection component, where the normalisation has been determined to remain approximately constant 
for each of the low-state spectra (see Section~\ref{sec:const}). 
This, along with the fact that the model component (\textsc{pexmon}) 
is not required to be absorbed by the variable neutral absorption that attenuates the intrinsic power law (also Section~\ref{sec:const}), 
suggests that this feature originates primarily from material that is exterior to the X-ray absorber. 
Considering that X-ray variability is often associated with material originating in the BLR region or inner torus, this evidence would then be in favour 
of an extended torus origin for the reflecting material, as expected by the the Unification model for AGN \citep{antonucci93}. 
Though this feature has been adequately accounted for by assuming constant reflection of the primary emission, some failure of the model is evident from positive residuals 
present at $\sim$\,6.4 keV (see Figure~\ref{fig:ratio}). It is possible that an additional contribution to the Fe K emission complex may originate from more 
internal regions of the torus that may partake in continuum absorption (see Section~\ref{sec:origin}). This would contaminate the extended (and therefore constant) 
Fe K emission region with emission from material that may be variable on shorter time scales. However, Fe K$\alpha$ emission from material originating at smaller distances from the 
continuum source (e.g. the accretion disk) is ruled out from the absence of a broad relativistic component in this feature (L09).

In addition to the power law component, characteristic of AGN emission, the soft X-ray spectra of this source show evidence for absorption by ionised material that 
is present in two phases (a feature also reported in A06 and L09). 
It has been shown that the warm absorber components are not observed to vary between Obs. 3 and 4 (see Section~\ref{sec:const} and Table~\ref{tab:warm_abs}).
While the exact origin of the warm absorbers observed in AGN is not well established, 
a study based on a sample of 23 AGN (Blustin et al. 2005) showed that this component is likely to originate in outﬂows from the circumnuclear torus 
(see also Kaastra et al. 2012), or even at kpc scale, as recently proposed by \citet{digesu13} for the source  1H 0419-577. 
Such an origin for the warm absorption in this source is tentatively supported by the lack of short term variability in the properties of the warm absorbers, 
since smaller distances from the SMBH would imply more rapid variability
(for recent examples see \citet{longinotti13} and \citet{gofford14}). 

As discussed in Section~\ref{sec:global model}, the global model required an emission component (\textsc{apec}) to account for residuals present in the soft X-ray band. 
Emission in the 0.5\,-\,2 keV spectrum is expected to occur from gas photoionised by the AGN \citep{bianchi06,guainazzi07}, 
as well as gas ionised by stellar activity in the host galaxy \citep{lamassa12}.
The low statistical quality of the RGS data makes it difficult to distinguish between these different processes, however, the following simple diagnostics can be used to 
determine which process is more likely.

First, if it is to be assumed that the luminosity of this component is entirely due to galactic star formation, an upper limit on the star formation rate (SFR) in the host 
galaxy can be determined via the relation SFR\,=\,2.2$\times$10$^{-40}$ L$_{0.5-2}$ M$_{\odot}$ yr$^{-1}$ \citep{ranalli03}. 
From the \textsc{apec} luminosity, L$^\text{APEC}_\text{0.5-2}$\,=\,1.36\,$\pm$\,${0.07}\times$10$^{41}$ erg s$^{-1}$, the estimate for the SFR is 
$\simeq$\,29.9\,$\pm$\,{1.5} M$_{\odot}$ yr$^{-1}$.
This SFR is unusually high and would classify H0557-385 as a starburst
galaxy, as these objects typically present SFRs of 5\,-\,50M$_{\odot}$\,yr$^\text{-1}$
within a region of 0.1\,-\,1 kpc extent \citep{kennicutt98}. As the optical
spectra do not show any sign of young stellar population compatible with a
starburst activity, these results suggest that it may not be appropriate
to attribute the emission of the X-ray component entirely
to star formation.

Following the method described in \citet{bianchi06}, the ratios of O\textsc{iii} to soft X-ray flux 
(using the flux of the APEC component, F$^\text{APEC}_\text{0.5\,-\,2}$\,=\,5.4\,$\pm$\,0.3$\times$10$^\text{-14}$ erg cm$^\text{-2}$ s$^\text{-1}$,
reported in Section~\ref{sec:global model}) for the optical observations 2010-10-17, 2010-11-03, 2011-01-30 were calculated to be 9.3, 9.6, and 6.4 respectively.     
These ratios are in the same range as those reported in \citet{bianchi06} for a sample of Seyfert 2 galaxies in which the soft X-ray emission regions have been shown to be
concident in spatial extent with the NLR, as mapped by the O\textsc{iii} emission. This evidence supports the view that the 0.5\,-\,2 keV emission in 
this source may originate from the photoionisation of NLR gas, rather than from star formation in the host galaxy.

\subsection{Physical Structure of the AGN H0557-385}\label{sec:structure}

In attempting to provide a physical interpretation of the spectral features observed in H0557-385, it is instructive to first establish a 
picture of the AGN structure based on its physical properties.
Using Equation 4 from \citet{zamfir10}, it is possible to derive an estimate for the black hole mass if 
the FWHM of the broad component of the H$\beta$ line and the luminosity of the continuum at 5100 \AA\ are known. 
Using the {\it{SOAR/Goodman}} optical spectrograph, the flux of this source at 5100 \AA\ was measured to be 
1.02$\pm0.03\times$10$^\text{-14}$erg s$^{-1}$ cm$^{-2}$ \AA$^{-1}$, 
which corresponds to a luminosity 
5100L$_{5100}$\,=\,1.31$\pm0.04\times$10$^\text{44}$erg s$^{-1}$.
From this result, and the FWHM of the broad component of the H$\beta$ line given in Section~\ref{sec:opt_spec}, 
the mass of the central black hole was found to be M$_{\rm BH}$\,$\sim$\,6.4$\times$10$^{7}$M${\rm \odot}$. 

The black hole mass can also be determined from the luminosity and FWHM of the broad component of the H$_{\rm \alpha}$ line, 
using Equation 9 from \citet{greene05}.
The flux of the broad component of the H$\alpha$ line was measured to be F$_{H\alpha_{BC}}$\,=\,1.30$\pm0.05\times$10$^\text{-12}$erg cm$^{-2}$ s$^{-1}$, 
giving a luminosity L$_{H\alpha_{BC}}$\,=\,3.23$\pm0.14\times$10$^\text{42}$ erg s$^{-1}$.
The FWHM was measured to be 4293 km s$^\text{-1}$.
From these results, it was found that the value for the black hole mass, M$_{\rm BH}$\,$\sim$\,6.4$\times$10$^{7}$M${\rm \odot}$,
agreed with the limits of Equation 9 from \citet{greene05}. Therefore, this value will be adopted for the remainder of this analysis.

Hereafter, a linear size of the X-ray source is assumed to be equal to D$_{s}$\,$\simeq$\,10R$_\text{g}$, where R$_\text{g}$\,=\,GM$_\text{BH}$/c$^{2}$,
as suggested for other AGN by micro-lensing \citep{chartas02,chartas09} and occultation \citep{risaliti07} experiments.
It is noted however, that the data presented here 
do not show the same ingress/egress detail that was available for NGC 1365. Using the value for the black hole mass given above gives D$_\text{s}$\,$\simeq$\,1$\times$10$^{14}$ cm.

Next, the dust sublimation radius will be estimated.
The dust sublimation radius was proposed as an upper boundary on the BLR \citep{netzer93},
interior to which dust grains are evaporated by emission from the central source.
This radius, which is often taken as representative of the inner boundary of the dusty torus, is given by

\begin{equation}\label{equ:dustsub}
 \mathrm{R_{d} \simeq 0.4\left(\frac{L}{10^{45} \, erg \,s^{-1}}\right)^{1/2}\left(\frac{1500\,K}{T_{sub}}\right)^{2.6} pc}
\end{equation}
 
\noindent \citep{nenkova08b} where L is the bolometric luminosity, and T$_\text{sub}$ is the dust grain evaporation temperature, generally taken to be the evaporation 
temperature of graphite grains, T\,$\sim$\,1500 K \citep{barvainis87}. It is noted here that the sharp boundary given by Equation~\ref{equ:dustsub} is an approximation. 
In reality the transition from dusty to dust-free regions is gradual, the radius at which dust grains can exist being dependent on the grain radii. Taking the maximum 
luminosity measured by the EPIC best-fit model (Obs. 4, Table~\ref{tab:multi_spec}) and applying a bolometric correction of 20 \citep{vasudevan07}) gives a 
bolometric luminosity of L$\rm_{BOL}$\,=\,2.5$\times$10$^{45}$ erg s$^{-1}$. Inserting this value into the above equation gives R$_\text{d}\simeq$ 2$\times$10$^{18}$ cm 
($\sim$ 0.6 pc). This result can also be used to estimate the outer edge of the torus, which is expected to lie in the range 
R$_\text{out}$\,$\simeq$\,5\,-\,30\,R$_\text{d}$ \citep[see][and reference therein]{miniutti14}. 
Taking the upper bound of R$_\text{out}$\,=\,30\,R$_\text{d}$ gives R$_\text{out}$\,$\simeq$\,5.5$\times$10$^{19}$ cm ($\sim$\,18 pc).

It is expected that the broad optical emission lines observed in AGN originate from material that is interior to the dust sublimation radius. From optical spectroscopy 
\citep{rodriguez00} three components of the broad H$\alpha$ emission line have been detected in H0557-385 with FWHM that correspond to 1035, 
2772, and 11000 km s$^{-1}$ in velocity space. Using the correction factor $\frac{\sqrt{3}}{2}$, which is appropriate when considering a spherical 
BLR \citep{zhang02}, the orbital velocities of the BLR clouds can be estimated from these line widths. The FWHM of the lines given above correspond to cloud 
velocities of 900, 2400, and 9550 km s$^{-1}$ respectively. 
Assuming that such clouds are in Keplerian motion around the source gives orbital radii of 1$\times$10$^{18}$, 1.5$\times$10$^{17}$, and 9$\times$10$^{15}$ cm respectively. 
As expected, these values are all within the upper limit set by the dust sublimation radius.

A second, independent estimate of the distance to the BLR can be made using the relationship between the radius of the BLR and the optical luminosity of the AGN. 
From Equation 1 in \citet[][see also \citet{bentz06}]{bentz07} the radius of the BLR can be found if the luminosity at 5100 \AA\ is known.  
From the continuum luminosity, 5100L$_{5100}$\,=\,1.31$\pm0.04\times$10$^\text{44}$erg s$^{-1}$,
the radius of the BLR was found to be R$_\text{BLR}\simeq$ 1$\times$10$^{17}$ cm, which is in agreement with the 
orbital radii of the BLR clouds determined above.

\begin{figure}
	\begin{center}
  	\psfig{figure=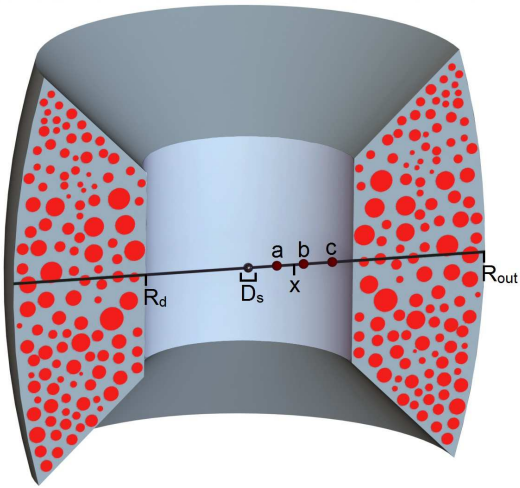,width=7cm,height=6.5cm}
	\caption{\label{fig:torus}
An illustration of the AGN geometry as described in Section~\ref{sec:structure}. The central black sphere represents the X-ray emission region, with linear size, 
D$_\text{s}$\,=\,1$\times$10$^{14}$ cm. The spheres labeled a, b, and c represent three components of the BLR at distances of 9$\times$10$^{15}$, 1.5$\times$10$^{17}$,
and 1$\times$10$^{18}$ cm respectively. The point labeled x represents the radius of the BLR (1$\times$10$^{17}$ cm) as inferred from both 
the 2\,-\,10 keV luminosity and the luminosity at 5100 \AA\ (see text for details).
Finally, the points R$_\text{d}$ and R$_{\text{out}}$ represent the estimated inner and outer radii of the circumnuclear torus, at distances of 2$\times$10$^{18}$
and 5.5$\times$10$^{19}$ cm respectively. The red circles illustrate the possible ``clumpy'' interior structure of the torus, as discussed in Section~\ref{sec:origin}. 
Distances are not to scale.}
 	\end{center}
\end{figure}

The distance to the BLR can be estimated in a similar way from Equation 1 in \citet{kaspi05} using the 2\,-\,10 keV luminosity. From this relation, and, as before, taking 
the maximum measured luminosity (Obs. 4, Table~\ref{tab:multi_spec}), the distance to the BLR was found to be R$_\text{BLR}$\,$\simeq$\,1$\times$10$^{17}$ cm. Again, 
this value is a good approximation to the previous estimations presented in this section. To allow for ease of comparison, the distance estimates that were derived in this 
section are illustrated in Figure~\ref{fig:torus}.

\subsection{Origin of the X-Ray Absorbers}\label{sec:origin}

In the last eight years, H0557-385 exhibited an X-ray spectrum typical of an obscured Seyfert 2 galaxy. The nuclear emission is seen through a partial covering cold absorber 
with an average column density of $\sim$7$\times$10$^{23}$~cm$^{-2}$. However, optical spectroscopy simultaneous with some of the X-ray observations in 2010 unveiled
a broad-line AGN, consistent with its historical classification as a Seyfert 1. Unless the discrepancy between the optical and the X-ray classification is due to 
different variability time scales, not adequately monitored by the sparse observations which are available, H0557-385 belongs to the class of AGN that do not fit the 
0th order Seyfert Unification scenario. Deep AGN surveys suggest that the fraction of these discrepant object could be as large as 30\% \citep{merloni14}.

There are two possible explanations for this behaviour: a) the X-ray partial covering clouds are located within the BLR; b) the X-ray absorbing clouds are dust-free 
(and can be located anywhere). 
We address in this Section the nature and location of the obscuring clouds in H0057-385, using time constraints derived from our monitoring campaigns.

The first evidence of dramatic X-ray absorption in this source is the transition from the intermediate-state (1995) to the earliest unabsorbed state (2000). 
The {\it{ASCA}} data (1995) suggest a moderate column density (N$_\text{H}$\,$\sim$\,2$\times$10$^{22}$ cm$^{-2}$) neutral absorber covering more than 90 per cent of the 
X-ray emitting region, while the {\it{BeppoSAX}} data (2000) show the source in an unabsorbed state. Assuming the simplest possible geometry, where a cloud moving 
with transverse velocity, V$_\text{c}$, with respect to our LOS moves a distance that is equal to at least the linear size of the X-ray source, D$_\text{s}$, in time 
$\Delta$T, gives the relation D$_\text{s}$\,$\leq$\,V$_\text{c}\Delta$T.

Rearranging the above equation to find the velocity of the cloud gives V$_\text{c}$\,$\geq$\,D$_\text{s}$/$\Delta$T, and for 
$\Delta$T\,=\,(T$_\text{Obs. 2}$\,-\,T$_\text{Obs. 1}$)\,=\,1.81$\times$10$^{8}$ s and the value for D$_\text{s}$ derived in Section~\ref{sec:structure}, 
the velocity of the cloud is V$_{c}$\,$\geq$\,5.5 km s$^{-1}$. 
Assuming the cloud is in Keplerian motion around the source, this velocity corresponds to a cloud orbiting at a radius R$_{c}$\,$\leq$\,3$\times$10$^{22}$ cm from the source. 
Following the same logic, but considering the obscuring event that occured sometime between Obs. 4 and 5 gives a 
cloud velocity V$_\text{c}$\,$\geq$\,8 km s$^{-1}$, 
refining the orbital radius 
to R$_\text{c}$\,$\leq$\,1$\times$10$^{22}$ cm. These extreme upper limits on the distance to the cloud arise from
the poor time constraints available on these events, and in principle 
they may be consistent with an absorber located in the dust lane or in the disc of the host Galaxy.

However, the X-ray absorbing clouds are not expected to exist
on such large scales, primarily because such an absorber
would also obscure the optical BLR, which is clearly observed in this
source (see Section~\ref{sec:opt_spec} and Figure~\ref{fig:H0557-opt}) during the X-ray obscured states of 2010. 
An absorber at these scales would also be expected to cause a drop in the UV flux, which again is not observed (Section~\ref{sec:opt_spec}). 
Further evidence of an inner absorber comes from the fact that in 2010 we observe weekly variation of the X-ray obscuration.  
To investigate the possibility that the absorber is located instead on much smaller scales, 
the occultation epoch from 2006 to 2011 is examined. It is assumed that the source remains covered by a single absorbing structure from Obs. 5 to the 
latest {\it{Swift}} observation, giving a minimum occultation time scale of $\Delta$T$_\text{occ,1}^\text{min}$\,=\,1.657$\times$10$^{8}$ s. 
The linear size of the cloud, D$_\text{c}$, can be related to V$_\text{c}$, $\Delta$T$_\text{occ}^\text{min}$, and D$_\text{s}$ according to the relation

\begin{equation}\label{equ:dc}
\mathrm{D_{c} \geq V_{c}\Delta T_{occ}^{min}+D_{s}}
\end{equation}
  
\begin{table*}
\centering
\caption[''long'']{\label{tab:cloud} The estimated properties of the X-ray absorber obtained by assuming that it either forms part of the BLR or the 
circumnuclear torus. An absorber column density of N$_\text{H}$\,=\,7.4$\times$10$^{23}$ cm$^{-2}$ is assumed (see text for details).}
 \begin{tabular}{l c c c c c}\\
\toprule
\midrule
\multicolumn{1}{c}{Velocity} &
\multicolumn{2}{c}{$\Delta$T$_\text{occ,1}^\text{min}$} &
\multicolumn{2}{c}{$\Delta$T$_\text{occ,2}^\text{min}$} \\

(km s$^{-1}$) & D$_\text{c}$ (cm) & N$_\text{e}$ (cm$^{-3}$) & D$_\text{c}$ (cm) & N$_\text{e}$ (cm$^{-3}$) \\
\midrule 
\multicolumn{5}{c}{Broad  Line  Region} \\

\midrule

900 & $\geq$ 1.5$\times$10$^{16}$ & $\leq$ 5$\times$10$^{7}$ & $\geq$ 4.7$\times$10$^{15}$ & $\leq$ 1.6$\times$10$^{8}$  \\

2400 & $\geq$ 4$\times$10$^{16}$ & $\leq$ 2$\times$10$^{7}$ & $\geq$ 1$\times$10$^{16}$ & $\leq$ 6$\times$10$^{7}$  \\

9550 & $\geq$ 1.6$\times$10$^{17}$ & $\leq$ 4.6$\times$10$^{6}$ & $\geq$ 5$\times$10$^{16}$ & $\leq$ 1.5$\times$10$^{7}$  \\

\midrule

\multicolumn{5}{c}{Torus} \\
\midrule

124 & $\geq$ 2$\times$10$^{15}$ & $\leq$ 4$\times$10$^{8}$ & $\geq$ 7$\times$10$^{14}$ & $\leq$ 1$\times$10$^{9}$ \\

650 & $\geq$ 1$\times$10$^{16}$ & $\leq$ 7$\times$10$^{7}$ & $\geq$ 3$\times$10$^{15}$ & $\leq$ 2$\times$10$^{8}$ \\

\midrule
\bottomrule

\end{tabular}
\end{table*}

\citep{miniutti14}. Inserting BLR cloud orbital velocities (see Section~\ref{sec:structure}) into Equation~\ref{equ:dc} yields the corresponding minimum cloud linear size 
that would be required to produce the observed obscuration. From these results, an estimate of the electron density, N$_\text{e}$, can be made from the relation 
N$_\text{e}$\,=\,N$_\text{H}$/D$_\text{c}$, where N$_\text{H}$ can be taken to be the average of the column densities measured in Obs. 5\,-\,9, giving 
N$_\text{H}$\,=\,7.4$\times$10$^{23}$ cm$^{-2}$. 
It is noted that during this occultation event, there is no information available on the absorption state of the source between Obs. 6 and 7, which corresponds to a 
time scale of $\sim$\,4 years, in which time the source may not necessarily remain in an obscured state. Therefore, the values for D$_\text{c}$ and N$_\text{e}$ were 
also calculated using a minimum occultation time scale that only extends from the start to the end of the {\it{Swift}} monitoring; 
$\Delta$T$_\text{occ,2}^\text{min}$\,=\,5.07$\times$10$^{7}$ s. 
As can be seen from Figure~\ref{fig:flux_swift}, the source is not observed to revert to an unobscured state during the course of this monitoring. 
However, it is noted that even the {\it{Swift}} campaign does not provide a continuous record of the flux of the source;
the largest interval between two observations being around three and a half months. 
Therefore, it is emphasised that all calculations involving $\Delta$T$_\text{occ,2}^\text{min}$ are based on the assumption that the source remains in an obscured state 
for the duration of the monitoring. The cloud properties obtained from these calculations are listed in Table~\ref{tab:cloud}.

An alternate interpretation is that the X-ray absorbers form part of the circumnuclear torus. 
As discussed in Section~\ref{sec:disc}, the inner and outer boundaries of the 
torus are given by R$_\text{d}$\,$\simeq$\,2$\times$10$^{18}$ and R$_\text{out}$\,$\simeq$\,5.5$\times$10$^{19}$ cm respectively. 
Assuming Keplerian motion, the velocities of material at these distances 
are V$_\text{d}$\,$\simeq$\,650 km s$^{-1}$ and V$_\text{out}$\,$\simeq$\,124 km s$^{-1}$.
Following the same formalism as before, the estimated cloud properties were calculated and are listed in Table~\ref{tab:cloud}.

The upper limits on the electron density, N$_\text{e}$, of the clouds at the BLR velocity using T$_\text{occ,1}$ are inconsistent with the expected density of 
the BLR \citep{popovic03}, and only marginally consistent if T$_\text{occ,2}$ is employed (see Table~\ref{tab:cloud}). On the other hand, the upper limits on the 
electron density of the clouds at the torus velocity are well consistent with the expected torus density \citep[see][]{elitzur06,miniutti14}. 
This evidence is considered an indication that the clouds responsible for the X-ray variability episodes 
in H0557-385 might be dust-free, and located beyond the BLR and within the dust sublimation radius.

As discussed in Section~\ref{sec:structure}, the dust sublimation radius does not represent a sharp boundary, but rather a 
region where gas gradually condenses into dust as distance from the source increases \citep{nenkova08b}. From this analysis, it is expected that, while the X-ray 
absorbing material may be located at the dust sublimation radius, it is likely to be primarily dust free. This is evident from the fact that broad optical emission
lines are clearly observed in this source (see Figure~\ref{fig:H0557-opt}) during X-ray obscured spectral states. An X-ray absorber located outside the BLR would be expected to obscure the 
broad emission lines, unless it is dust-free. The fact that broad optical emission lines are observed in this source is strong evidence in favour
of a dust-free absorber.

Considering then, that the X-ray absorber must be dust free, it is suggested here that it lies in the intervening region, exterior to the BLR, but interior to the inner 
boundary of the ``dusty'' torus. 
This deduction gives credence to the suggestion that the BLR and torus may in fact be two 
components of a single, continuous medium that surrounds the nucleus, toroidal in shape, where the dust-to-gas ratio gradually increases in proportion to the distance
from the central continuum source, a scenario that has been observationally supported via infrared spectroscopy 
by the evidence that the outer radius of the BLR is limited by dust \citep{landt14}.

Furthermore, in order to fully account for the X-ray spectral variability observed in this source, it must be assumed that the X-ray absorbing medium is not 
homogeneous, but rather consists of discrete ``clumps'' (or clouds) of gas, as illustrated in Figure~\ref{fig:torus}. This would explain the observed variation among the 
low-state spectra, and the subsequent variations in the covering fraction of the absorber, as measured by the spectral model defined in Section~\ref{sec:global model}. 
A clumpy torus absorber would allow intrinsic emission from the AGN to leak through, and give rise to variations in the soft X-ray spectrum, even among obscured states. 
This phenomenon is clearly observed in H0557-385. 
Finally, a clumpy absorber would imply that the probability of directly observing the (unobscured) AGN continuum would be finite, and dependent on the individual cloud 
size and number density. In light of this interpretation, the high-state (unobscured) spectra ({\it{XMM-Newton}} 2002, {\it{BeppoSAX}} 2000) would represent epochs when 
the observers LOS did not intercept material in the clumpy environment which caused the obscuration event(s) of 2006/2010.

\section{Conclusions}\label{sect:Conclusion}

In this analysis, the X-ray spectral variability associated with the Seyfert 1 AGN H0557-385 has been interpreted as being due to the occultation of the central emission 
region by high column density (N$_\text{H}$\,$\sim$\,7$\times$10$^{23}$ cm$^{-2}$) neutral material, covering $\ge$\,80 per cent of the X-ray emission region.

From consideration of the absorption time scales, it has been inferred that the absorbing material forms part of the clumpy circumnuclear torus, at a distance 
of $\sim$\,2$\times$10$^{18}$ cm from the X-ray emission region. 
The detection of broad optical emission lines in this source implies that the obscuration occurs 
in a dust-free region within, or very close to, the dust sublimation radius. 
It may be possible that the LOS grazes the diffuse upper atmosphere of the clumpy torus, which would explain why a Compton-thick absorption state, 
normally associated with obscuration by the circumnuclear torus, is never observed in this source.

In order to further understand the nature of the absorbing medium in this source, it would be necessary to first observe an ``unveiling'' event, that is, to monitor 
the emersion of the central continuum source from the absorbed state. If such observations were available, it may be possible to explore the evolution of the cloud 
covering fraction with time, and hence place tighter constraints on the geometry and location of the obscuring clouds. In addition, observations of this source at 
energies $>$\,10 keV would be useful in attempting to determine the ionisation state of the absorbing medium (see Section~\ref{sec:variable_spec}). This data would 
then provide an additional estimate of the distance to the absorbing cloud based on the measured value of the ionisation parameter. 
This measurement could be possible with deep {\it{NuSTAR}} or {\it{Astro-H}} observations.

\section*{Acknowledgments}

This work is based on data obtained with {\it{XMM-Newton}}, an ESA science mission with instruments and contributions directly funded by ESA Member States and NASA. 
Data was also provided by the Tartarus Database (version 3.1), operated by the Tartarus Team (Imperial College London). 
We thank Rogerio Riffel for guidance and useful discussions on the use of the STARLIGHT code.
D.C. acknowledges fincancial support from the Faculty of the European Space Astronomy Centre (ESAC) as well as from the Enterprise Ireland Space Education Programme. 
A.L.L. acknowledges support by NASA contract numbers NNX10AK91G and NNX12AE83G.
Support for this work was provided by the National Aeronautics and Space Administration through the Smithsonian Astrophysical Observatory contract SV3-73016 to MIT 
for support of the HETG project.
The research leading to these results has received funding from the European Commission Seventh Framework Programme (FP7/2007-2013) 
under grant agreement n.267251 ``Astronomy Fellowships in Italy'' (AstroFIt).
ARA thanks to Conselho National de Desenvolvimento Cient\'{i}fico e
Tecnol\'{o}gico (CNPq) for partial support of this work through grant
307403/2012-2.

\bibliographystyle{mn2e}
\bibliography{bibliography}

\label{lastpage}

\end{document}